%% file: main.tex
\documentclass[journal]{IEEEtran}
\usepackage{times,graphicx,amssymb,amsmath,algorithm, algorithmic,subfigure,cite,url}

\newtheorem{prop}{Proposition}[section]

\newtheorem{assum}{Assumption}[section]
\newtheorem{rem}{Remark}[section]
\newtheorem{alg}{Algorithm}[section]

\newcommand{\argmin}{\operatornamewithlimits{argmin}}

\newcommand{\RR}{\mathbb{R}}

\newcommand{\calI}{\mathcal{I}}
\newcommand{\calJ}{\mathcal{J}}

\newcommand{\calO}{\mathcal{O}}

\newcommand{\calX}{\mathcal{X}}
\newcommand{\calY}{\mathcal{Y}}

\title{An Alternating Direction Method Approach
to \\Cloud Traffic Management}
\author{Chen~Feng,~\IEEEmembership{Member,~IEEE,} Hong~Xu,~\IEEEmembership{Member,~IEEE,}
and~Baochun~Li,~\IEEEmembership{Senior~Member,~IEEE}}

\begin{document}
\maketitle

\begin{abstract}
In this paper, we introduce a unified framework for studying various cloud traffic management problems,
ranging from geographical load balancing to backbone traffic engineering. 
We first abstract these real-world problems as a multi-facility resource allocation problem,
and then present two distributed optimization algorithms by exploiting the special structure
of the problem. Our algorithms are inspired by Alternating Direction Method of Multipliers (ADMM), 
enjoying a number of unique features. 
Compared to dual decomposition, 
they converge with non-strictly convex objective functions;
compared to other ADMM-type algorithms, they not only achieve 
faster convergence under weaker assumptions,
but also have lower computational complexity
and lower message-passing overhead. 
The simulation results not only confirm these desirable features of
our algorithms, but also highlight
several additional advantages, such as scalability 
and fault-tolerance.
\end{abstract}

\input{intro}
\input{application-new}

\input{admm}
\input{simulation}
\input{related}
\input{conclusion}

\bibliographystyle{abbrv}
\bibliography{IEEEabrv,main}
\input{appendix}

\end{document}

%% file: intro.tex
\section{Introduction}
\label{sec:intro}

Cloud services (such as search, social networking, etc.) 
are often deployed on a geographically distributed
infrastructure, i.e., data centers located in different
regions. In order to optimize the efficiency of these data centers, 
how to orchestrate the data transmission, 
including traffic flowing from users to the infrastructure 
to access the cloud services, and traffic flowing across these 
data centers for back-end services, has started to receive
an increasing amount of attention. We refer to these problems generally as 
\emph{cloud traffic management} herein.

In this paper, we introduce a unified framework for 
studying various cloud traffic management problems,
ranging from geographical load balancing to backbone traffic engineering. 
As we will see in Sec.~\ref{sec:app}, a large variety of
cloud traffic management problems can be abstracted into
the following form:
\begin{align}\label{opt:form1}
\text{maximize} &\quad \sum_{i = 1}^N f_i(x_{i1}, \ldots, x_{in}) - \sum_{j=1}^n g_j(y_j) \\
\text{subject to} &\quad \forall j: \sum_{i = 1}^N x_{ij} = y_j \nonumber \\
 &\quad \forall i: x_i=(x_{i1}, \ldots, x_{in})^T \in \calX_i \subseteq \RR^n \nonumber\\
&\quad \forall j: y_j \in \calY_j \subseteq \RR. \nonumber
\end{align}
Generically, the problem \eqref{opt:form1}
amounts to allocating resources from $n$ facilities to $N$ users 
such that the ``social welfare'' (i.e., utility minus cost) is maximized. 
The utility function
$f_i(x_i)$ represents the performance, or the level of satisfaction,
of user $i$ when she receives an amount $x_{ij}$ of resources from
each facility $j$, where $x_i = (x_{i1}, \ldots, x_{in})^T$.
In practice, this performance measure can be in terms of revenue,
throughput, or average latency, depending on the problem setup.
We assume throughout the paper that $f_i(\cdot)$ are concave.
The cost function $g_j(y_j)$ represents the operational
expense or congestion cost when facility $j$ allocates an 
amount $y_j$ of resources to all the users. Note that $y_j$ is the 
sum of $x_{ij}$ (over $i$), since each facility often
cares about the total amount of allocated resources. 
We assume that $g_j(\cdot)$ are convex.
The constraint sets $\{\calX_i\}$ and $\{\calY_j\}$ are used to model
the additional constraints, which are assumed to be 
compact convex sets.

We refer to problem \eqref{opt:form1} as the \emph{multi-facility
resource allocation problem}. We are particularly interested
in solutions that are amenable to \emph{parallel} implementations,
since a cloud provider usually has abundant servers for parallel computing.
For a production cloud, \eqref{opt:form1} is inherently a large-scale
convex optimization problem, with millions of variables, or even more.
The standard approach to constructing parallel algorithms
is dual decomposition 
with (sub)gradient methods. However, it suffers from several
difficulties for problem \eqref{opt:form1}. 
First, dual decomposition requires a delicate adjustment 
of the step-size parameters, which have a strong influence on the convergence
rate.  
Second, dual decomposition requires the utility functions
$f_i(\cdot)$ to be strictly concave \emph{and} the cost functions $g_j(\cdot)$ to be
strictly convex to achieve convergence.
These requirements cannot be met in many problem 
settings of \eqref{opt:form1} we will demonstrate in Sec.~\ref{sec:app}.

To overcome these difficulties, we develop a new decomposition
method for the multi-facility resource allocation problem. 
Unlike dual decomposition, our method uses a \emph{single} parameter, which is much easier to tune
and often leads to fast convergence.
More importantly, our method converges with 
\emph{non-strictly} concave utility functions and 
\emph{non-strictly} convex cost functions.

Our decomposition method is based on
\emph{alternating direction method of multipliers} (ADMM), 
a simple yet powerful method that has recently found 
practical use in many large-scale convex
optimization problems \cite{BPCP10}. Although ADMM has been widely applied to areas of machine learning and signal processing, 
its application to networking research is still in an early stage.
To the best of our knowledge, our previous work \cite{XL13, Henry-sig, icac13} represents one of the first such applications.
Very recently,  interesting applications to radio-access networks and fuel cell generation in geo-distributed cloud
 have been proposed in \cite{LHFLLZ14} and \cite{ZLLLJZL14}, respectively.

Compared to these previous algorithms, the algorithms developed in this 
paper require much weaker technical assumptions to ensure convergence, 
and, at the same time,
enjoy much lower computational complexity and message-passing overhead. 

Finally, we present an extensive empirical study of our algorithms. 
Our simulation results reveal some additional advantages of our algorithms, 
including their scalability to a large number of users and
their fault-tolerance with respect to updating failures.

The main contributions of this paper are as follows:
\begin{enumerate}
\item We identify several cloud traffic management problems
as instances of the multi-facility resource allocation problem \eqref{opt:form1}.
 
\item We develop two distributed algorithms for problem
\eqref{opt:form1}, which enjoy a number of unique advantages
over dual decomposition and previous ADMM-based algorithms.

\item We present extensive simulation results, which further demonstrate
the scalability and fault-tolerance of our algorithms.
\end{enumerate}

%% file: application-new.tex
\section{Applications to Cloud Traffic Management} \label{sec:app}

In this section, we will show that a large variety of
optimization problems in the context of cloud traffic management are indeed 
instances of the multi-facility resource allocation problem \eqref{opt:form1}.
In particular, we will illustrate the  inherent large scale of these problems for production systems,
and explain why the utility function is non-strictly concave and the cost function is non-strictly
convex for some applications.

\subsection{Geographical Load Balancing}

\subsubsection{Background}

Cloud services, such as search, social networking, etc., 
are often deployed on a geographically distributed infrastructure, i.e. data centers located in different regions as shown in Fig. 1, for better performance and reliability. A natural question is then how to direct the workload from users among the set of geo-distributed data centers in order to achieve a desired trade-off between performance and cost, since the energy price exhibits a significant degree of geographical diversity as seminally pointed out by \cite{QWBG09}. This question has attracted much attention recently \cite{LCBW12,icac13,Henry-sig,QWBG09,LLWL11,XL13,GCWK12}, and is generally referred to as geographical load balancing.

\begin{figure}[htp]
\centering
\includegraphics[width=0.47\textwidth]{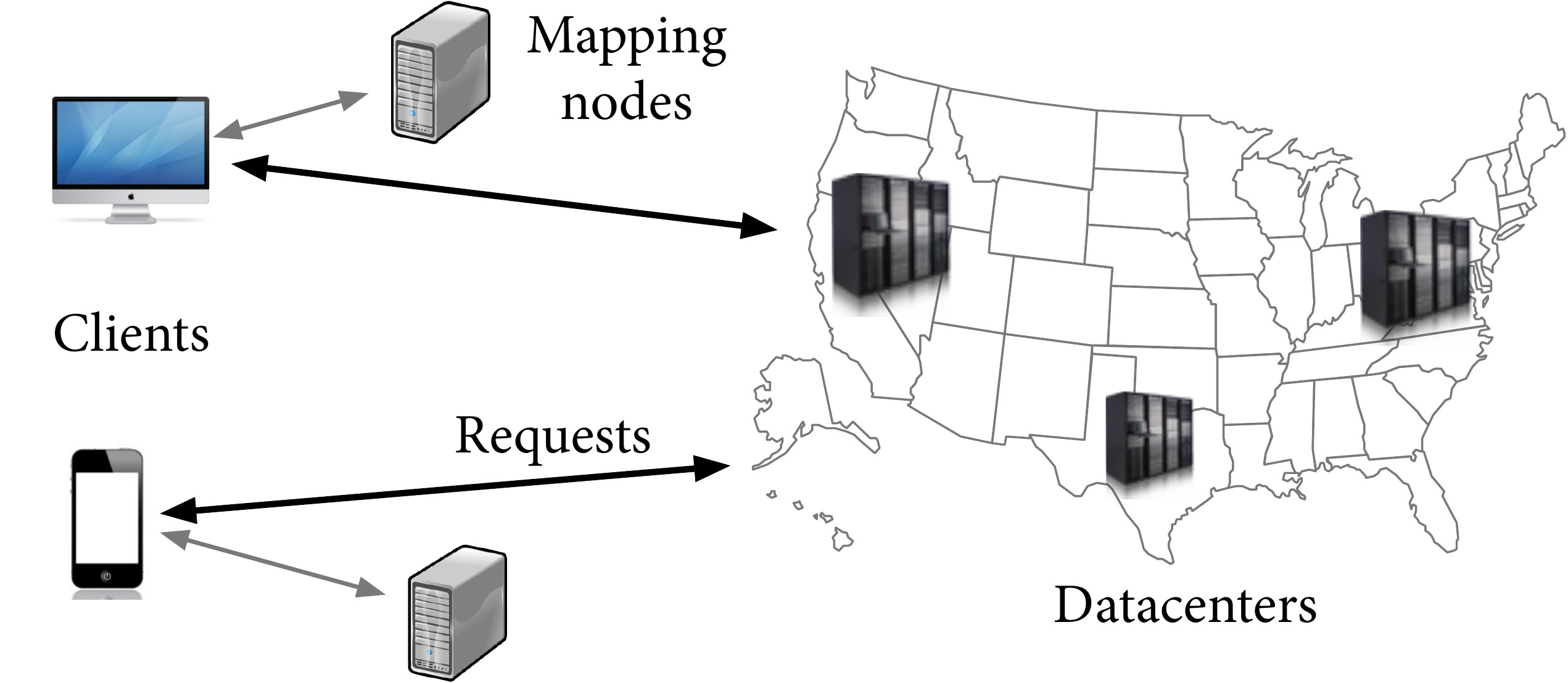}
\caption{A cloud service running on geographically distributed data centers.}
\label{fig:showcase}
\end{figure}

\subsubsection{Basic Model}

We now introduce a formulation for the basic geographical load balancing problem, which captures the essential performance-cost trade-off and covers many existing works \cite{icac13,Henry-sig,QWBG09,LLWL11,XL13,GCWK12}.  
Here, we define a user to be an group of customers aggregated  
from a common geographical region sharing a unique IP prefix, 
as is often done in practice  to reduce complexity \cite{NSS10}. 
We use $x_{ij}$ to denote the amount of workload coming from
user $i$ and directed to data center $j$. We use 
$t_i$ to denote the total workload of each user. 
We use $f_i(\cdot)$ to represent the utility of user $i$,
and use $g_j(\cdot)$ to represent the cost of data center $j$.
These functions can take various forms depending on the scenario as we will elaborate soon.

With these notations, we formulate the basic geographical load balancing problem:
\begin{align} 
	\text{maximize} &\quad \sum_{i} f_i(x_{i}) - \sum_{j} g_j\left(y_j\right) \label{opt:basic} \\
	\text{subject to} &\quad \forall i: \ \sum_{j} x_{ij} = t_i,
	x_i \in \RR_{+}^n, \label{con:conservation} \\
	&\quad \forall j: \ y_j = \sum_{i} x_{ij} \le c_j, \label{con:capacity}
\end{align}
where  
\eqref{con:conservation} describes the workload conservation 
and non-negativity constraint, and \eqref{con:capacity} is the capacity constraint at data centers.
Since the constraint \eqref{con:conservation} can be rewritten as
$\forall i: x_i \in \calX_i$, where $\calX_i$ is a convex
set, problem \eqref{opt:basic} is an instance of problem \eqref{opt:form1}.

Now, let us consider the utility function $f_i(\cdot)$.
Latency is arguably the most important performance metric for most interactive services: A small increase in the user-perceived latency can cause substantial utility loss for the users \cite{KHS07}. The user-perceived latency largely 
depends on the end-to-end propagation latency \cite{FMLC03,NJRC12},
which can be obtained through active measurements.
Let $l_{ij}$ denote the end-to-end propagation latency between user~$i$ and 
data center~$j$. The following utility function $f_i$ has been used in \cite{Henry-sig,icac13}
\begin{equation}
\label{eqn:f_glb}
f_i(x_i) = -q t_i \left( \sum_j x_{ij} l_{ij}/t_i \right)^2.
\end{equation}
Here, $q$ is the weight factor that captures the relative importance of performance compared to cost in monetary terms. Clearly, the utility function $f_i(\cdot)$
achieves its maximum value when latency is zero. Also, the function $f_i(\cdot)$
depends on the \emph{average latency}
$\sum_j x_{ij} l_{ij}/t_i$. For different applications, $f_i$ may depend on other aggregate statistics of the latency, such as the maximum latency or the 99-th percentile latency, which may be modeled after a norm function.

For the cost function $g_j(\cdot)$, many existing works consider the following \cite{QWBG09,LLWL11,XL13,GCWK12}
\begin{equation}
\label{eqn:g_glb}
g_j(y_j) = P^E_j\cdot \text{PUE} \cdot E(y_j).
\end{equation}
Here, $P^E_j$ denotes the energy price in terms of \$/KWh at data center $j$. PUE, power usage effectiveness, is the ratio between total infrastructure power and server power. Since total infrastructure power mainly consists of server power and cooling power, PUE is commonly used as a measure of data center energy efficiency. Finally, $E(y_j)$ represents the server power at data center~$j$, which is a function of the total workload $y_j$ and can be obtained empirically. A commonly used server power function is from a measurement study of Google \cite{FWB07}:
\begin{equation}\label{eqn:dcpower}
	E(y_j) = c_j P_{\text{idle}}+\left( P_{\text{peak}} - P_{\text{idle}} \right)y_j,
\end{equation}
where $P_{\text{idle}}$ is server idle power and $P_{\text{peak}}$ peak power.

\subsubsection{Problem Scale}

The geographical load balancing problem \eqref{opt:basic} would be easy to solve, if its scale is small with, say, hundreds of variables. However, for a production cloud, \eqref{opt:basic} is inherently an extremely large-scale optimization. In practice, the number of users $N$ (unique IP prefixes) is on the order of $\calO(10^5)$ \cite{NSS10}. Thus the number of variables $\{x_{ij}\}$ is $\calO(10^6)$. The load balancing decision usually needs to be updated on a hourly basis, or even more frequently, as demand varies dynamically. The conventional dual decomposition approach suffers from many performance issues for solving such large-scale problems, as we argued in Sec.~\ref{sec:intro}. Thus we are motivated to consider new distributed optimization algorithms. 

\subsubsection{Extensions}

In this section, we provide some additional extensions of the basic model \eqref{opt:basic} from the literature to demonstrate its importance and generality.

\textbf{Minimizing Carbon Footprint.}
In \eqref{opt:basic}, the monetary cost of energy is modeled. The environmental cost of energy, i.e., the carbon footprint of energy can also be taken into account. Carbon footprint also has geographical diversity due to different sources of electricity generation in different locations \cite{GCWK12}. Hence, it can be readily modeled by having an additional carbon cost $P^C_j$ in terms of average carbon emission per KWh in the objective function of \eqref{opt:basic} following \cite{GCWK12,LLWL11}. 


\textbf{Joint Optimization with Batch Workloads.}
There are also efforts \cite{LCBW12,icac13,Henry-sig} that consider the delay-tolerant batch workloads in addition to interactive requests, and the integrated workload management problem. Examples of batch workloads include MapReduce jobs, data mining tasks, etc. Batch workloads provides additional flexibility for geographical load balancing: Since their resource allocation is elastic, when the demand spikes we can allocate more capacity to run interactive workloads by reducing the resources for batch workloads. 

To incorporate batch workloads, we introduce
$n$ ``virtual'' users, where user $j$ generates batch workloads
running on
data center $j$. Let $w_j$ be the amount of resource used for batch workloads on data center $j$,
and let $\tilde{f}_j(w_j)$ be the utility of these batch
workloads. Then the joint optimization can be formulated as follows:
\begin{align*}
	\text{maximize} &\quad \sum_{i}f_i(x_i) + \sum_{j} \tilde{f}_j(w_j) - \sum_{j} g_j(y_j)\\
	\text{subject to} &\quad \forall i: \ \sum_j x_{ij} = t_i, \ x_i \in \RR_{+}^n; w \in \RR_{+}^n\\
	&\quad \forall j: \ y_j = \sum_{i} x_{ij} +  w_j \le c_j. \nonumber
\end{align*}
The utility function $\tilde{f}_j(\cdot)$ depends only on $w_j$ but not on latency, due to its elastic nature. In general, 
$\tilde{f}_j(\cdot)$ is an increasing and concave function, such as the log function used in \cite{icac13,Henry-sig}. 
Clearly, this is still an instance of \eqref{opt:form1}.

\subsection{Backbone Traffic Engineering}

\subsubsection{Background}

Large cloud service providers, such as Google and Microsoft, usually interconnect 
their geo-distributed data centers with a private backbone wide-area networks (WANs). 
Compared to ISP WANs, data center backbone WANs exhibit unique characteristics \cite{JKMO13,GHCR13}.
First, they are increasingly taking advantage of the software-defined networking (SDN) architecture, where a logically centralized controller has global knowledge and coordinates all transmissions \cite{GHMM05,CCFR05}.
SDN paves the way for implementing logically centralized traffic engineering. 
In addition, the majority of the backbone traffic, such as copying user data to remote 
data centers and synchronizing large data sets across data centers, is elastic.
Thus, since the cloud service provider controls both the applications at the edge
and the routers in the network, in addition to routing, it can perform application rate control, i.e., allocate the aggregated sending rate of each application, according to the current network state. 
These characteristics open up the opportunity to perform joint rate control and
traffic engineering in backbone WANs, which is starting to receive attention in the networking community \cite{JKMO13,HKMZ13,GHCR13}.

\subsubsection{Basic Model} 

We model the backbone WAN as a set $\calJ$ of interconnecting links. 
Conceptually, each cloud application generates a \emph{flow} between a source-destination pair of data centers. 
We index the flows by $i$, and denote by $\calI$ the set of all flows.
We assume that each flow can use multiple paths from its source to destination. This is because
multi-path routing is relatively easy to implement (e.g., using MPLS \cite{HKMZ13,JKMO13,EJLW01}) and offers many benefits.
For each flow $i$, we denote by $\mathcal{P}_i$ the set of its available paths
and define a \emph{topology matrix} $A_i$ of size $|\mathcal{J}| \times |\mathcal{P}_i|$
as follows:
\[
A_i[j, p] = \begin{cases}
1, &\mbox{if link } j \mbox{ lies on path } p \\   
0, &\mbox{otherwise.}
\end{cases}
\]
\vspace{-4mm}
\begin{figure}[htp]
\centering
\includegraphics[width=0.26\textwidth]{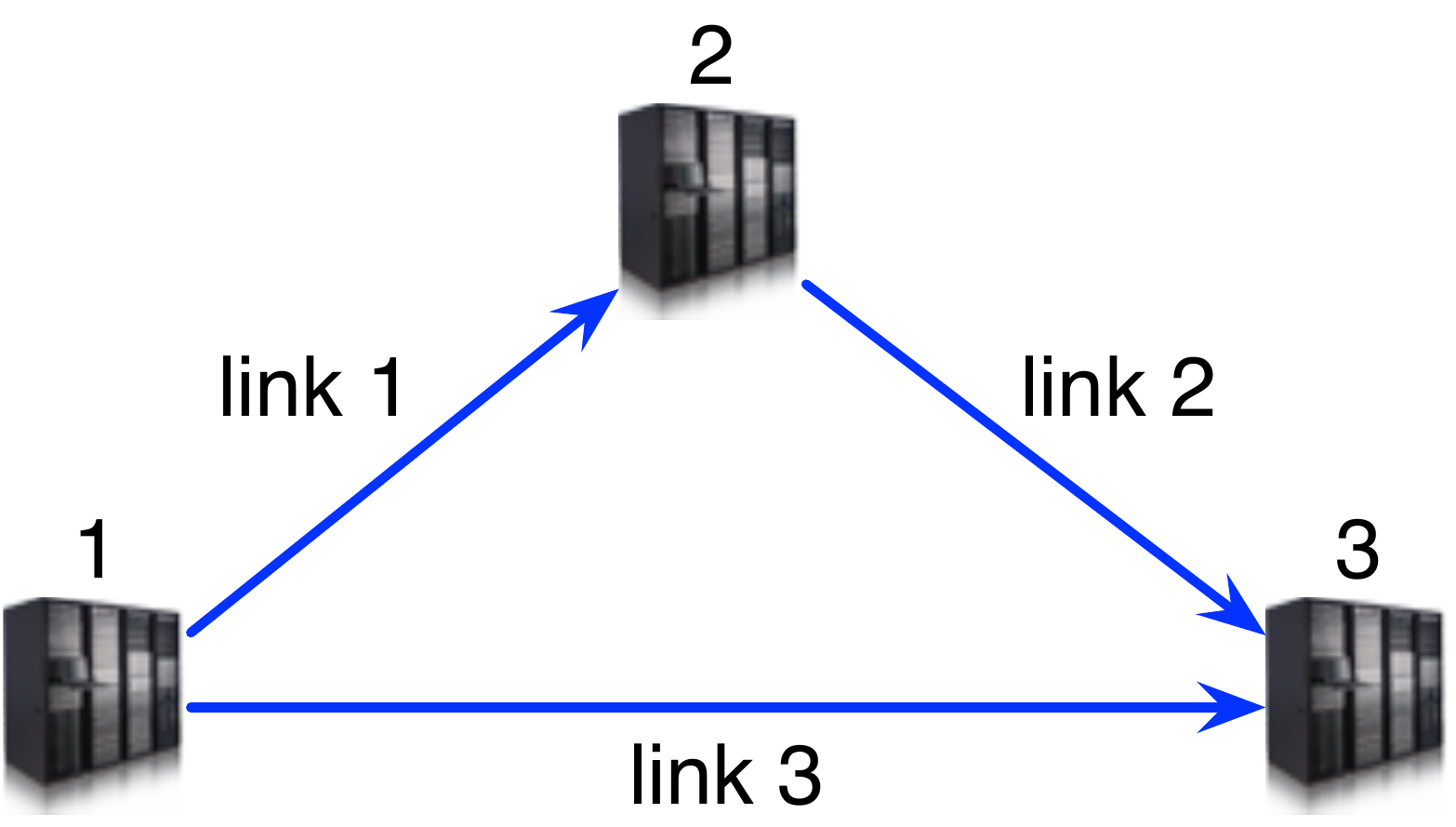}
\caption{An illustration of three data centers with 3 links.}
\label{fig:illustration}
\end{figure}

For example, consider a network with three data centers 
and 3 links as illustrated in Fig.~\ref{fig:illustration}. A flow (say, flow $1$) from data center~1 to data center~3 has two paths:
$\{ \mbox{link 1}, \mbox{link 2}  \}$ and $\{\mbox{link 3}\}$.
In this case, $|\calJ| = 3$, $|\mathcal{P}_1| = 2$, and the topology matrix $A_1$ is
\[
A_1 = \begin{bmatrix}
1 & 0\\1 & 0\\0 &1
\end{bmatrix}.
\]
Clearly, the topology matrix $A_i$ provides a mapping
from paths to links. Let $w_{ip}$ denote the amount of traffic of
flow $i$ on path $p$, and let $x_{ij}$ denote the amount of traffic
of flow $i$ on link $j$. Then we have $x_i = A_i w_i$,
where $w_i = (w_{i1}, \ldots, w_{i|\mathcal{P}_i|})^T$.
Since $A_i$ is always full column-rank (otherwise some path
must be redundant), $A_i$ has a left-inverse $A_i^{-1}$ such that
$w_i = A_i^{-1} x_i$. For instance, a left-inverse 
of $A_1$ in the previous example is
\[
A_1^{-1} = \begin{bmatrix}
1 & 0 & 0\\0 & 0 &1
\end{bmatrix}.
\]

Note that $w_i$ models the rate control decision for each application flow. A flow corresponds to potentially many TCP connections between a particular source-destination pair of data centers, carrying traffic for this particular application. We choose to model rate control at the application flow level because the latest data center backbone architectures \cite{HKMZ13,JKMO13} are designed to control the aggregated sending rates of applications across data centers. The aggregated rate can be readily apportioned among different connections following some notion of fairness, and rate control can be enforced by adding a shim layer in the servers' operating system and using a per-destination token bucket \cite{BCKR11a}. 

We use $f_i(w_i)$ to represent the utility of flow $i$, and
$g_j(y_j)$ to represent the congestion cost of link $j$,
where $y_j = \sum_i x_{ij}$ is the total traffic on link $j$.
The joint rate control and traffic engineering problem can be
formulated as
\begin{align} 
	\text{maximize} &\quad \sum_{i} f_i(A_i^{-1}x_{i}) - \sum_{j} g_j\left(y_j\right) \label{opt:basic2} \\
	\text{subject to} &\quad \forall i: \ 
	x_i \in \RR_{+}^n, \label{con:non-negative2} \\
	&\quad \forall j: \ y_j = \sum_{i} x_{ij} \le c_j, \label{con:capacity2}
\end{align}
where \eqref{con:non-negative2} describes the non-negativity
constraint, and \eqref{con:capacity2} says that the total
traffic on link $j$ cannot exceed the capacity $c_j$.
Clearly, problem \eqref{opt:basic2} is again an instance
of problem \eqref{opt:form1}.

The utility function $f_i(w_i)$ should be concave,
such as the $\log$ function
$f_i(w_i) = \log(\sum_p w_{ip})$, or a more 
general ``rate-fairness'' function used for
Internet TCP congestion control \cite{MW00}.
It is worth noting that even if $f_i(w_i)$ is strictly
concave (with respect to $w_i$), $f_i(A_i^{-1} x_i)$ 
is \emph{not} strictly concave (with respect to $x_i$) 
in general. This important fact has been used in Sec.~\ref{sec:compare} to demonstrate the advantages
of our distributed algorithms.
The cost function $g_j(y_j)$ is convex and non-decreasing.
For example, the function can be a piece-wise linear
function with increasing slopes, which is used in \cite{GHCR13}.

Finally, note that the topology matrix $A_i$ only depends on 
the source-destination pair. Hence, for a given source data center,
the number of all possible topology matrices is bounded by
the number of all other data centers, which is typically less than $30$. 
In other words, the topology matrices are easy to store and maintain in practice.
Note also that all the inverse matrices can be computed before the algorithm runs.
That is, there is no need to calculate any $A_i^{-1}$ on the fly.

\subsubsection{Problem Scale}

Similar to the geographical load balancing problem, backbone traffic engineering
is also a large-scale optimization problem for a production data center backbone WAN. 
In practice, a provider runs hundreds to thousands of applications with around ten data centers \cite{HKMZ13,JKMO13}. Thus the number of application flows is $\calO(10^5)$ to $\calO(10^6)$. For a WAN with tens of links, we potentially have tens of millions of variables $\{ x_{ij} \}$. Compared to geographical load balancing, the traffic engineering decisions need to be updated over a very small time window (say, every 5 or 10 minutes as in \cite{HKMZ13,JKMO13}) to cope with traffic dynamics. This further motivates us to derive a fast distributed solution. 

\subsubsection{Extensions} We present some possible
extensions of the basic model.

\textbf{Minimizing Bandwidth Costs.} 
Unlike big players like Google and Microsoft, small cloud providers often rely on ISPs to interconnect their
data centers. In this case, bandwidth costs become one of the most important  
operating expenses. Although many ISPs adopt the 95-percentile
charging scheme in reality, the link bandwidth cost is often 
assumed to be linear with the link traffic, 
because optimizing a linear cost in each interval can 
reduce the monthly 95-percentile bill \cite{ZZGH10}. 
Hence, the bandwidth cost can be easily incorporated by adding
these linear functions to \eqref{opt:basic2}.

\textbf{Incrementally Deployed SDN.}
Instead of upgrading all routers to be SDN-capable with a daunting bill, 
cloud providers could deploy SDN incrementally \cite{AKL13}. 
In such a scenario, some routers still use standard routing
protocols such as OSPF, while other routers have the flexibility 
to choose the next hop. This scenario can be easily handled 
by imposing additional constraints on the set $\mathcal{P}_i$ of 
available paths such that $\mathcal{P}_i$ only contains 
\emph{admissible} paths. (See Definition~1 in \cite{AKL13} for details.)
Clearly, with some routers restricted to standard protocols, 
the number $|\mathcal{P}_i|$ of available paths for flow $i$ is reduced, 
resulting in a smaller-scale optimization problem.

%% file: admm.tex
\section{ADMM-Based Distributed Algorithms}\label{sec:algorithm}

In this section, we will present two ADMM-based distributed algorithms
that are well suited for the multi-facility resource allocation problem,
with a particular focus on their convergence rates 
as well as their advantages over other ADMM-based algorithms.

\subsection{A Primer on ADMM}

We begin with some basics of ADMM and its connection to
dual decomposition. Dual decomposition is a standard approach to solving 
large-scale convex problems, which has been widely used
in the networking research. By forming the Lagrangian
for problem \eqref{opt:form1} (with the Lagrange multiplies
$\lambda \in \RR^n$) and applying the dual decomposition
method, one arrives at the following algorithm:

\begin{alg}\label{alg:dual}
Initialize $\{x_{i}^0 \}$,
$\{y_j^0 \}$,  $\{ \lambda_j^0 \}$. For $k=0,1,\ldots,$
 repeat
\begin{enumerate}
	\item {\bf $x$-update:} Each user $i$ solves the following sub-problem for $x_i^{k+1}$:
\begin{align*}
	\text{min} &\quad  -f_i(x_i) + (\lambda^k)^T x_i\\
	\text{s.t. } &\quad x_i \in \calX_i. 
\end{align*}
	\item {\bf $y$-update:} Each facility $j$ solves the following
	sub-problem for $y_j^{k+1}$:
\begin{align*}
	\text{min} &\quad g_j(y_j) - \lambda_j^k y_j\\
	\text{s.t. } &\quad y_j \in \calY_j.
\end{align*}
	\item {\bf Dual update:} Each facility $j$ updates $\lambda_j^{k+1}$:
\begin{equation*}
\lambda_j^{k+1} := \lambda_j^k + \rho^k \left(\sum_{i=1}^N x_{ij}^{k+1} - y_j^{k+1} \right),
\end{equation*}
where $\rho^k$ is the step-size for the $k$th iteration.
\end{enumerate}
\end{alg}

The following assumption is valid throughout the paper:
\begin{assum}\label{ass:ass1}
The optimal solution set of problem \eqref{opt:form1} is non-empty, 
and the optimal value $p^*$ is finite.
\end{assum}

It is known that Algorithm~\ref{alg:dual} is convergent
under Assumption~\ref{ass:ass1} and the assumption that
the utility functions $f_i(\cdot)$ are strictly concave
and the cost functions $g_j(\cdot)$ are strictly convex
\cite{BT97}. However, as we have shown in Sec.~\ref{sec:app},
for many interesting problems
of form \eqref{opt:form1}, either $f_i(\cdot)$ are
\emph{non-strictly} concave or $g_j(\cdot)$ are 
\emph{non-strictly} convex, making conventional dual decomposition
unsuitable for such applications.

Alternating direction method of multipliers (ADMM) is a 
decomposition method that does not require strict convexity.
It solves convex optimization problems in the form
\begin{align}\label{opt:admm}
	\text{minimize} &\quad f(x) + g(y)\\
	\text{subject to} &\quad Ax + By = c, \nonumber \\
				&\quad x \in \calX, \ y \in \calY, \nonumber
\end{align}
with variables $x \in \RR^n$ and $y \in \RR^m$, where
$f: \RR^n \to \RR$ and $g: \RR^m \to \RR$ are convex functions,
$A \in \RR^{p\times n}$ and $B \in \RR^{p \times m}$ are matrices,
$\calX$ and $\calY$ are nonempty compact convex subsets of
$\RR^n$ and $\RR^m$, respectively. 
Note that $f(\cdot)$ and/or $g(\cdot)$ are \emph{not} assumed 
to be strictly convex.
  
The \emph{augmented Lagrangian} \cite{H69} for problem \eqref{opt:admm} is
\begin{multline*}
 	L_\rho(x, y, \lambda) = f(x) + g(y) +\lambda^T(Ax + By - c)\\+(\rho/2)\|Ax + By - c\|^2_2,
\end{multline*}
where $\lambda \in \RR^p$ is the Lagrange multiplier 
(or the dual variable) for
the equality constraint, and  $\rho>0$ is the 
\emph{penalty parameter}.
Clearly, $L_0$ is the (standard) Lagrangian for \eqref{opt:admm},
and $L_\rho$ is the sum of $L_0$ and
a \emph{penalty term} $(\rho/2)\|Ax + By - c\|^2_2$.
 
The standard ADMM algorithm solves problem \eqref{opt:admm} with the iterations \cite{BPCP10}:
\begin{align*}
	x^{k+1} &:= \argmin_{x\in \calX} L_\rho(x, y^k, \lambda^k), \\
	y^{k+1} &:= \argmin_{y \in \calY} L_\rho(x^{k+1}, y,
	\lambda^k), \\
 	\lambda^{k+1} &:= \lambda^k + \rho( A x^{k+1} + B y^{k+1} - c),
\end{align*}
where the penalty parameter $\rho$ can be viewed as the step size for the update of the dual variable $\lambda$. 
Note that the primal variables $x$ and $y$ are updated in an
alternating fashion, which accounts for the term
\emph{alternating direction}. 

The standard ADMM algorithm has a \emph{scaled form}, which is often
more convenient (and will be used in this paper). 
Introducing $u = (1/\rho) \lambda$ and
combining the linear and quadratic terms in the augmented
Lagrangian, we can express the ADMM algorithm as
 \begin{align*}
 	x^{k+1} &:= \argmin_{x\in \calX} \left( f(x) + 
 	(\rho/2) \| Ax + B y^k - c + u^k \|_2^2 \right), \\
 	y^{k+1} &:= \argmin_{y \in \calY} \left( g(y) +
 	(\rho/2) \| A x^{k+1} + B y - c + u^k \|_2^2 \right), \\
  	u^{k+1} &:= u^k + A x^{k+1} + B y^{k+1} - c.
 \end{align*}
 
Applying this algorithm to problem \eqref{opt:form1}, we
obtain the following algorithm:
\begin{alg}\label{alg:admm}
Initialize $\{x_{i}^0 \}$,
$\{y_j^0 \}$,  $\{ u_j^0 \}$. For $k=0,1,\ldots,$
 repeat
\begin{enumerate}
	\item {\bf $x$-update:} The users jointly solve the following problem for $\{ x_i^{k+1} \}$:
\begin{align*}
	\text{min} &\quad  -\sum_{i=1}^N f_i(x_i) + 
	(\rho/2) \| \sum_i x_i - y^k + u^k \|_2^2\\
	\text{s.t. } &\quad \forall i: x_i \in \calX_i. 
\end{align*}
	\item {\bf $y$-update:} Each facility $j$ solves the following
	sub-problem for $y_j^{k+1}$:
\begin{align*}
	\text{min} &\quad g_j(y_j) + (\rho/2)
	\left( \sum_{i=1}^N x_{ij}^{k+1} - y_j + u_j^k \right)^2\\
	\text{s.t. } &\quad y_j \in \calY_j.
\end{align*}
	\item {\bf Dual update:} Each facility $j$ updates $u_j^{k+1}$:
\begin{equation*}
u_j^{k+1} := u_j^k + \sum_{i=1}^N x_{ij}^{k+1} - y_j^{k+1}.
\end{equation*}
\end{enumerate}
\end{alg}

It is known that Algorithm~\ref{alg:admm} is convergent
under Assumption~\ref{ass:ass1} \cite{BT97}. 
However, the $x$-update
requires all the users to solve a joint optimization 
due to the penalty term $(\rho/2) \| \sum_i x_i - y^k + u^k \|_2^2$, which is undesirable for large-scale systems.

\subsection{Distributed ADMM Algorithms}

Here, we present two distributed ADMM algorithms, 
as well as their convergence analysis. The first algorithm
is essentially the same as the ADMM-based method for the
sharing problem \cite[Chapter~7]{BPCP10}. 
The second algorithm is a variation of the first one
by switching the order of $x$-update and $y$-update.

\begin{alg}\cite[Chapter~7]{BPCP10}\label{alg:admm1}
Initialize $\{x_{i}^0 \}$,
$\{y_j^0 \}$,  $\{ u_j^0 \}$. For $k=0,1,\ldots,$
 repeat
\begin{enumerate}
	\item {\bf $x$-update:} Each user $i$ solves the following sub-problem for $x_i^{k+1}$:
\begin{align*}
	\text{min} &\quad  -f_i(x_i) + (\rho/2) \| x_i - x_i^k
	+ d^k  \|_2^2\\
	\text{s.t. } &\quad x_i \in \calX_i,
\end{align*}
where $d^k \triangleq (1/N)\left( u^k + \sum_{i=1}^N x_{i}^k
- y^k \right)$.
	\item {\bf $y$-update:} Each facility $j$ solves the following
	sub-problem for $y_j^{k+1}$:
\begin{align*}
	\text{min} &\quad g_j(y_j) + (\rho/2N) 
	\left(y_j  - \sum_{i=1}^N {x}_{ij}^{k+1} -	u_j^k \right)^2\\
	\text{s.t. } &\quad y_j \in \calY_j.
\end{align*}
	\item {\bf Dual update:} Each facility $j$ updates $u_j^{k+1}$:
\begin{equation*}
u_j^{k+1} := {u}_j^k + \sum_{i=1}^N x_{ij}^{k+1} - y_{j}^{k+1}.
\end{equation*}
\end{enumerate}
\end{alg}

\begin{alg}\label{alg:admm2}
Initialize $\{x_{i}^0 \}$,
$\{y_j^0 \}$,  $\{ u_j^0 \}$. For $k=0,1,\ldots,$
 repeat
\begin{enumerate}
	\item {\bf $y$-update:} Each facility $j$ solves the following
	sub-problem for $y_j^{k+1}$:
\begin{align*}
	\text{min} &\quad g_j(y_j) + (\rho/2N) \left(y_j  - \sum_{i=1}^N {x}_{ij}^k -	u_j^k \right)^2\\
	\text{s.t. } &\quad y_j \in \calY_j.
\end{align*}
	\item {\bf $x$-update:} Each user $i$ solves the following sub-problem for $x_i^{k+1}$:
\begin{align*}
	\text{min} &\quad  -f_i(x_i) + (\rho/2) \| x_i - x_i^k
	+ d^k \|_2^2\\
	\text{s.t. } &\quad x_i \in \calX_i,
\end{align*}
where $d^k \triangleq (1/N) \left( u^k + \sum_{i=1}^N x_{i}^k
- y^{k+1} \right)$.
	\item {\bf Dual update:} Each facility $j$ updates $u_j^{k+1}$:
\begin{equation*}
u_j^{k+1} := u_j^k + \sum_{i=1}^N x_{ij}^{k+1} - y_{j}^{k+1}.
\end{equation*}
\end{enumerate}
\end{alg}

Clearly, both algorithms preserve the separability of the problem. Moreover,
both algorithms have the same number of dual variables as dual decomposition and
the standard ADMM algorithm.
The convergence of Algorithm~\ref{alg:admm1} is established in \cite[Chapter~7]{BPCP10}
by showing that it is a variant of the standard ADMM algorithm. 
The convergence of Algorithm~\ref{alg:admm2} follows immediately from the convergence of
Algorithm~\ref{alg:admm1}. The connection between these two algorithms 
and the standard ADMM is provided in the Appendix.

More interestingly, by combining the above convergence result with several
very recent results \cite{HeYuan12,HeY12,DY12} on ADMM, we can characterize the
convergence rates of Algorithms~\ref{alg:admm1} and \ref{alg:admm2},
which are absent in  \cite[Chapter~7]{BPCP10}.
It turns out that both algorithms have an $\calO(1/k)$ rate of convergence
for the general case.
Moreover, if the cost functions $g_j(\cdot)$ are strictly
convex and their gradients $\nabla g_j(\cdot)$ 
are Lipschitz continuous, Algorithm~\ref{alg:admm1} achieves
linear convergence, i.e., convergence at rate $\calO(1/a^k)$
for some $a > 1$. Similarly, if the utility functions
$f_i(\cdot)$ are strictly concave and their 
gradients $\nabla f_i(\cdot)$ are Lipschitz continuous,
Algorithm~\ref{alg:admm2} achieves linear convergence.
Hence, Algorithms~\ref{alg:admm1} and \ref{alg:admm2}
naturally complement each other, as summarized in Table~\ref{table:twoalgs}.

Finally, we elaborate on the convergence rates presented in
Table~\ref{table:twoalgs},
since some of these results will be useful for our simulations. Without
loss of generality, we only focus on Algorithm~\ref{alg:admm1}.

Let $\left(\{x_i^*\}, \{z_i^*\}  \right)$ be a primal optimal
solution to problem \eqref{opt:form3} (in particular, we have $x_i^* = z_i^*$), 
and $\{\lambda_i^*\}$ be a dual optimal solution. Let $v_i^* = \lambda_i^*/\rho$. (The existence of $\{\lambda_i^*\}$ follows from
the strong duality theorem.)

\begin{prop}\label{thm:sublinear1}
Let $\{ \{x_i^k\}, y^k, u^k \}$ be any sequence 
generated by Algorithm~\ref{alg:admm1}.
Let $v^k = u^k / N$ and $z_i^k = x_i^k + v^{k-1} - v^k$. 
Let 
\begin{equation}\label{eq:Vk1}
V^k = \sum_{i = 1}^N \left( \|z_i^k - z_i^*\|_2^2 + 
\| v^k - v_i^*\|_2^2 \right),
\end{equation}
and
\begin{equation}\label{eq:Dk1}
D^k = \sum_{i = 1}^N \left( \|z_i^{k+1} - z_i^k\|_2^2 + 
\| v^{k+1} - v^k\|_2^2 \right).
\end{equation}
Then starting with any initial 
point $\{ \{x_i^0\}, y^0, u^0 \}$, $D^k$ is non-increasing, and  $D^k
\le {V^0}/{(k+1)}$ for all $k$.
\end{prop}

\begin{table}[tp]
\centering
\caption{Convergence rates of Algorithms~\ref{alg:admm1}
and \ref{alg:admm2}.}
\vspace{0.5em}
\small
{
\renewcommand{\arraystretch}{1.2}
\begin{tabular}{|c|c|c|c|c|}
\hline
 &   & Lipschitz  & Recommended & \\
\raisebox{0.6em}{Case} & \raisebox{0.6em}{Strictly convex} & continuous & {algorithms} & \raisebox{0.6em}{Rate}\\
\hline \hline
1 & none & none & Alg.~\ref{alg:admm1} or \ref{alg:admm2} & $\calO(1/k)$\\ \hline
2 & $\{ g_j\}$ & $\{ \nabla g_j\}$ & Alg.~\ref{alg:admm1} & $\calO(1/a^k)$\\ \hline
3 & $\{-f_i\}$ & $\{\nabla f_i\}$ & Alg.~\ref{alg:admm2} & $\calO(1/a^k)$\\ \hline
4 & $\{-f_i\}$, $\{ g_j\}$ & $\{\nabla f_i\}$, $\{ \nabla g_j\}$ & Alg.~\ref{alg:admm1} or \ref{alg:admm2} & $\calO(1/a^k)$\\ \hline
\end{tabular}
}
\label{table:twoalgs}
\end{table}

\begin{rem}
Proposition~\ref{thm:sublinear1}  suggests that 
the sequence $\{D^k\}$ 
can be used as a natural
stopping rule for Algorithm~\ref{alg:admm1}, which
decreases at rate $1/k$. This stopping rule is more
rigorous compared to that in \cite[Chapter~7]{BPCP10}, since
their stopping rule is based on heuristic principles.
For example, their stopping-rule sequence does not have the
non-increasing property and may fluctuate over iterations.
\end{rem}

\begin{prop}\label{thm:linear1}
Let $\{ \{x_i^k\}, y^k, u^k \}$ be any sequence 
generated by Algorithm~\ref{alg:admm1}. 
Let $V^k$ be the Lyapunov function defined in \eqref{eq:Vk1}.
Assume that the cost functions $g_j(\cdot)$ are 
strictly convex with Lipschitz continuous gradients.
Then starting with any initial 
point $\{ \{x_i^0\}, y^0, u^0 \}$,
there exists some $\delta > 0$ such that
$V^k \le V^0/ (1+ \delta)^k$ for all $k$.
\end{prop}

\begin{rem}
Proposition~\ref{thm:linear1} 
provides a guideline for choosing the 
penalty parameter $\rho$.
In particular, one can show that the parameter
$\delta = \min\{ c_9/\rho, c_{11} \rho \}$,
where $c_9$ and $c_{11}$ are given in \cite{DY12}. 
Hence, $\rho$ can be chosen such that
the parameter $\delta$ is maximized.
\end{rem}

The proofs of Propositions~\ref{thm:linear1} and \ref{thm:sublinear1}
are slight modifications of those presented in 
\cite{HeYuan12,HeY12,DY12}\footnote{We do not provide
the proofs here, but could include them upon editor's
request.}. Note that both algorithms use a \emph{single} parameter $\rho$,
which is easier to tune compared to dual decomposition. This is desirable
for parallel implementation.

\subsection{Parallel Implementation}
Here, we discuss how the above two algorithms can be effectively
implemented on parallel processors in a cloud environment.
with a particular focus on Algorithm~\ref{alg:admm1}, since the same discussion applies to Algorithm~\ref{alg:admm2}.

We associate each user a type-1 processor, which stores and maintains two states 
$(x_i^k, d^k)$. Similarly, we associate each facility
a type-2 processor, which stores and maintains $(u_j^k, \sum_i x_{ij}^{k+1})$.
At the $k$-th iteration, each type-1 processor solves a small-scale convex problem
(in $n$ variables), and then reports the updated $x_{ij}^{k+1}$ to facility $j$.
Each facility $j$ collects $x_{ij}^{k+1}$ from all type-1 processors, and then
computes the sum $\sum_i x_{ij}^{k+1}$. This is called a \emph{reduce} step
in parallel computing \cite{DG04}. After the reduce step, each type-2 processor solves
a single-variable convex problem for $y_j^{k+1}$ and updates $u_j^{k+1}$. Then, each type-2 
processor sends the value of $d_j^{k+1} \triangleq
(1/N) \left( u_j^{k+1} + \sum_i x_{ij}^{k+1} - y_j^{k+1} \right)$ to
all type-1 processors, which is called a \emph{broadcast} step.
Hence, each iteration consists of a reduce step and a broadcast
step, performing message-passing between different types of processors.


An alternative and perhaps simpler method to implement Algorithm~\ref{alg:admm1} is based on
the MPI \emph{Allreduce} operation \cite{MPI96}, which computes the global sum over
all processors and distributes the result to every processor. 
Although the Allreduce operation can be achieved by a reduce step followed 
by a broadcast step, an efficient implementation
(for example, via butterfly mixing) often leads to much better performance.
With the help of Allreduce, we only need $N$ processors
of the same type, with each storing and maintaining three states
$(x_i^k, u^k, \sum_i x_i^k)$. At the $k$-th iteration, each processor solves a small
convex problem and updates $x_i^{k+1}$. Then, all the processors perform an
Allreduce operation  so that all of them
(redundantly) obtain $\sum_i x_i^{k+1}$. After this Allreduce step, each processor 
solves $n$ single-variable convex problems and (redundantly) computes $u^{k+1}$. Clearly, this method simplifies the implementation and can potentially increase the speed.

\subsection{Comparisons with Other Algorithms}\label{sec:compare}

In this section, we compare Algorithms~\ref{alg:admm1} and \ref{alg:admm2} with dual decomposition and other ADMM-based
algorithms.

Algorithm~\ref{alg:dual} is the dual-decomposition algorithm  for problem \eqref{opt:form1}. Clearly, at each iteration,
it has essentially the same complexity as Algorithms~\ref{alg:admm1} and \ref{alg:admm2}.
However, Algorithm~\ref{alg:dual} 
requires delicate adjustments of
step sizes $\rho^k$, often resulting in slow convergence.
For instance, as we will show in Sec.~\ref{sec:sim_dual},
for solving the geographical load balancing problem \eqref{opt:basic},
Algorithm~\ref{alg:dual} does not converge after hundreds of iterations with a diminishing step-size rule \cite{BM}, while Algorithm~\ref{alg:admm1}
converges after 50 iterations.
Moreover, Algorithm~\ref{alg:dual} requires the cost functions
$g_j(\cdot)$ to be strictly convex and the utility functions
$f_i(\cdot)$ to be strictly concave to ensure convergence.
In contrast, Algorithms~\ref{alg:admm1} and \ref{alg:admm2}
do not make these assumptions.

There are some other ADMM-type distributed algorithms in the
literature, such as linearized ADMM \cite{HeY12} and 
multi-block ADMM \cite{HY12, HL12}. However, they are not
particularly suitable for the multi-facility resource allocation 
problem \eqref{opt:form1}. For example, applying linearized
ADMM to problem \eqref{opt:form1} gives the following iterations:
\begin{align*}
x_i^{k+1} &:= \argmin_{x_i \in \calX_i} \left( -f_i(x_i) +  x_i^T g^k
 + (r/2) \| x_i - x_i^k \|_2^2 \right)\\
y_j^{k+1} &:= \argmin_{y_j \in \calY_j} \left( g_j(y_j) + (\rho/2)
( y_j - \sum_{i = 1}^N x_{ij}^{k+1} - u_j^k  )^2 \right)\\
u_j^{k+1} &:= u_j^k + \sum_{i = 1}^N x_{ij}^{k+1} - y_j^{k+1},
\end{align*}
where $g^k = \rho(\sum_{i} x_i^k - y^{k} + u^k)$ \emph{linearizes} 
the penalty term $(\rho/2) \| \sum_{i} x_i - y\|_2^2$,
and $(r/2) \| x_i - x_i^k \|_2^2$ is a \emph{proximal term}.
Although the above algorithm preserves separability of the
problem,
its convergence requires $r > \rho N$.
When $N$ is sufficiently large, the $x$-update in each iteration
just slightly changes $x_i$ (due to a large $r$), making the convergence slow.
Hence, linearized ADMM is not well suited for large-scale problems.

Multi-block ADMM is another candidate for solving
problem \eqref{opt:form1}. However, it generally requires 
users to solve their subproblems sequentially rather than in parallel. 
Moreover, it still lacks theoretical
convergence guarantees for the general case. 
Indeed, a counter-example has recently been reported showing the impossibility
of convergence of multi-block ADMM for the general
case \cite{CHYY13}.

The algorithms presented in \cite{WHML13} are most similar
to ours. Their basic idea is also to apply variants of
the standard ADMM 
algorithm to solve separable convex problems.
However, their algorithms 
require the utility functions to be strictly concave
\emph{and} the cost functions to be strictly convex in order to achieve $\calO(1/a^k)$ rate of convergence. 
Such requirements cannot be met in some application
scenarios. One such example is backbone traffic engineering,
as we will discuss in Sec.~\ref{sec:app}.


Compared to our previous ADMM-based algorithms \cite{XL13,Henry-sig,icac13}, our algorithms proposed in
this paper enjoy a number of advantages. 
First, they assume weaker technical assumptions to ensure convergence.
Second, they have lower computational
complexity and lower message-passing overhead.
For example, the algorithms in \cite{icac13,Henry-sig}
require strictly convex objective functions
and bounded level sets to achieve convergence.
In contrast, our new algorithms converge with \emph{non-strictly}
convex objective functions.
As another example, the algorithm in \cite{XL13}
needs each datacenter (facility) to solve a large-scale 
quadratic problem at each iteration, whereas
our new algorithms only require each facility
to solve a single-variable convex problem 
at each iteration.

%% file: simulation.tex
\section{Empirical Study}\label{sec:simulation}

We present our empirical study of the performance of the distributed ADMM algorithms. For this purpose, it suffices to choose one of the two cloud traffic management problems since they are equivalent in nature. We use the geographical load balancing problem \eqref{opt:basic} with the utility and cost functions \eqref{eqn:f_glb} and \eqref{eqn:g_glb} as the concrete context of the performance evaluation. This problem corresponds to the most general case (i.e., case~1 in Table~\ref{table:twoalgs}), since \eqref{eqn:f_glb} is non-strictly concave and 
\eqref{eqn:g_glb} is non-strictly convex. Thus it can be solved using either Algorithm~\ref{alg:admm1} or Algorithm~\ref{alg:admm2}. We use Algorithm~\ref{alg:admm1} in all of our simulations. Note that if the objective function exhibits strict convexity, better simulation results can be obtained according to Proposition~\ref{thm:linear1}. In other words, we mainly focus on the ``worse-case'' performance of the algorithms in this section. We plan to make all our simulation codes publicly available after the review cycle.

\subsection{Setup}
\label{sec:setup}

We randomly generate each user's request demand $t_i$, with an average of $9\times 10^4$. We then normalize the workloads to the number of servers, assuming each request requires 10\% of a server's CPU. We assume the prediction of request demand is done accurately since prediction error is immaterial to performance of the optimization algorithms. The latency $l_{ij}$ between an arbitrary pair of user and data center is randomly generated between 50~ms and 100~ms. 

We set the number of data centers (facilities) $n=10$. Each data center's capacity $c_j$ is randomly generated so that the total capacity $\sum_j c_j$ is 1.4x the total demand. We use the 2011 annual average day-ahead on peak prices \cite{F11} at 10 different local markets as the power prices $P_j$ for data centers. The servers have peak power $P_{\text{peak}}=200$~W, and consume 50\% power at idle. The PUE is 1.5. These numbers represent state-of-the-art data center hardware \cite{FWB07,QWBG09}.

We set the penalty parameter $\rho$ of the ADMM algorithm to $\rho=10^{-3}$ after an empirical sweep of 
$\rho \in \{ 10^{-4}, 10^{-3}, \ldots, 10^{3}, 10^{4} \}$. 
Although a more fine-grained search for $\rho$ can further improve the performance of our algorithms, we confine ourselves to the above $9$ choices to demonstrate the practicality.

\begin{figure*}[htbp]
	\begin{minipage}[ht]{0.33\linewidth}
	\centering
	\includegraphics[width=1\linewidth]{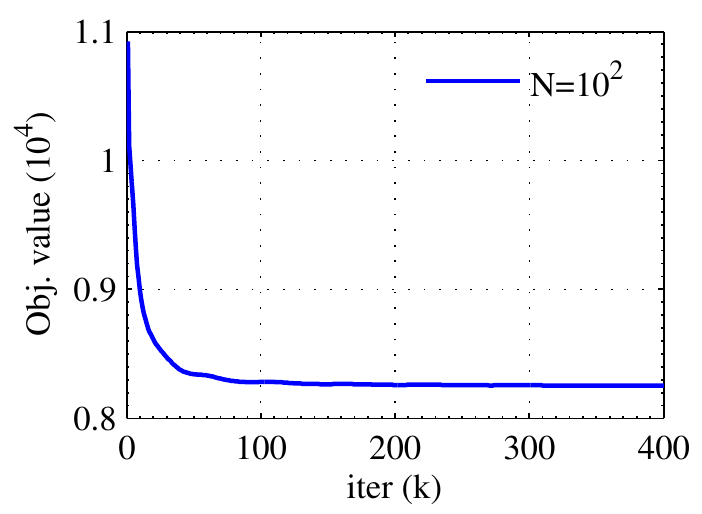}
	\caption{Objective value. $N=10^2$.}
	\label{fig:obj}
	\end{minipage}
	\begin{minipage}[ht]{0.33\linewidth}
	\centering
	\includegraphics[width=1\linewidth]{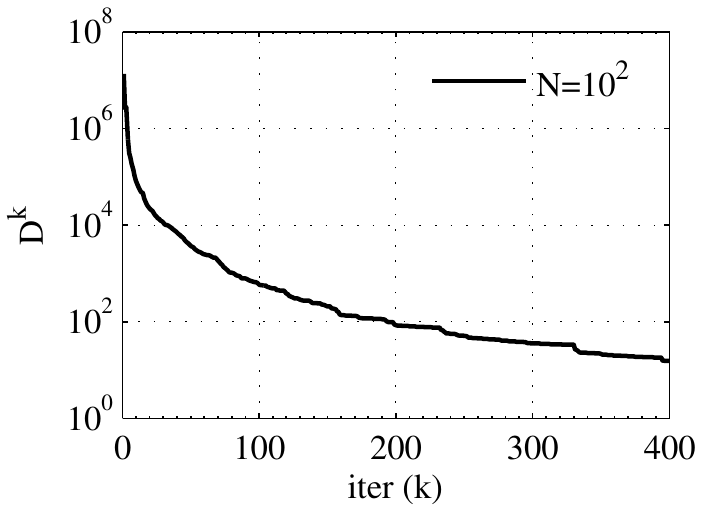}
	\caption{$D^k$. $N=10^2$.}
	\label{fig:Dk}
	\end{minipage}
	\begin{minipage}[ht]{0.33\linewidth}
	\centering
	\includegraphics[width=1\linewidth]{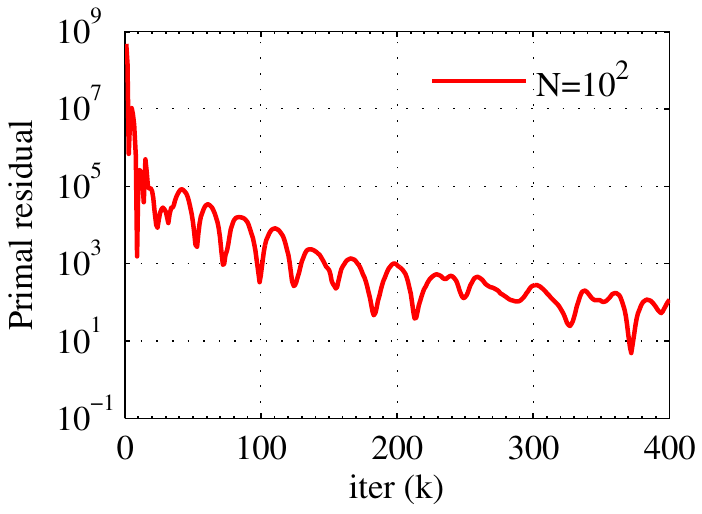}
	\caption{Primal residual. $N=10^2$.}
	\label{fig:primal}
	\end{minipage}
\end{figure*}
\begin{figure*}[htbp]
	\begin{minipage}[ht]{0.33\linewidth}
	\centering
	\includegraphics[width=1\linewidth]{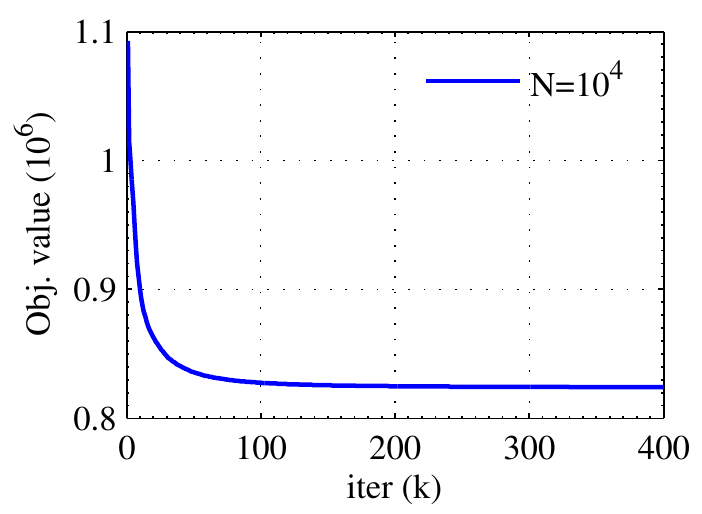}
	\caption{Objective value. $N=10^4$.}
	\label{fig:obj-1e4}
	\end{minipage}
	\begin{minipage}[ht]{0.33\linewidth}
	\centering
	\includegraphics[width=1\linewidth]{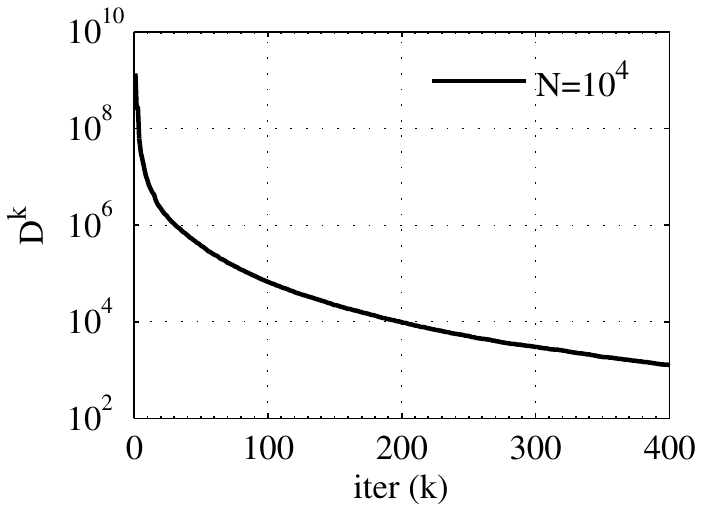}
	\caption{$D^k$. $N=10^4$.}
	\label{fig:Dk-1e4}
	\end{minipage}
	\begin{minipage}[ht]{0.33\linewidth}
	\centering
	\includegraphics[width=1\linewidth]{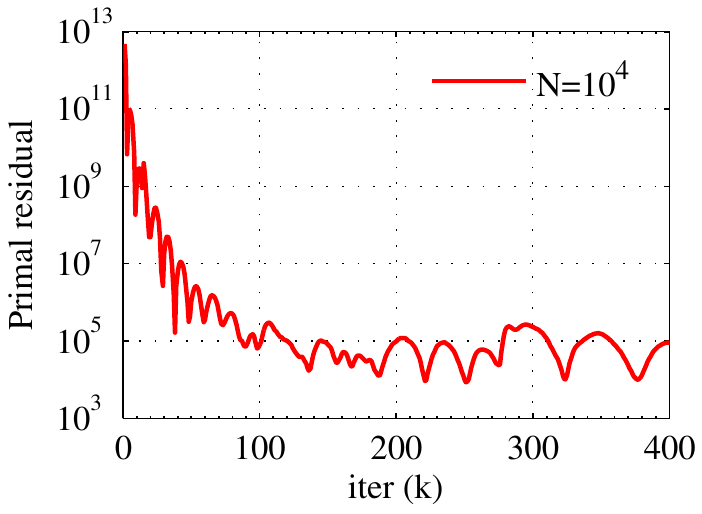}
	\caption{Primal residual. $N=10^4$.}
	\label{fig:primal-1e4}
	\end{minipage}
\end{figure*}

\subsection{Convergence and Scalability}

We evaluate the convergence of Algorithm~1 under the previous setup.
We vary the problem size
by changing the number of users $N \in \{ 10^2,10^3,10^4,10^5 \}$ 
and scaling data center capacities linearly with $N$.
We observe that our algorithm converges quickly after $50$
iterations in all cases, \emph{independent} of the problem size.

{\bf Convergence of objective functions.}
Figure~\ref{fig:obj} and \ref{fig:obj-1e4} plot the 
convergence of objective values for $N=10^2$ and $N = 10^4$, respectively.
Notice that the objective values for $N = 10^4$ are roughly $100$
times the corresponding values for $N = 10^2$ at each iteration. 
This means that our algorithm has excellent scalability,
which is very helpful in practice. Since the number of iterations is independent of the problem size, it suggests that our algorithm can solve a large-scale problem with (almost) the same running time by simply scaling the amount of computing resources linearly with the number of users.

{\bf Convergence of $D^k$.}
Figure~\ref{fig:Dk} and \ref{fig:Dk-1e4} show the trajectory of $D^k$ as defined in \eqref{eq:Dk1} for $N=10^2$ and $N = 10^4$, respectively.
We observe that $D^k$ is indeed non-increasing in both cases. Further, the two figures are in log scale, implying that $D^k$ decreases sublinearly,
which confirms Proposition~\ref{thm:sublinear1} for the $\calO(1/k)$ convergence rate.
In addition, one can see that $D^k$ scales linearly with $N$ as expected from its definition. This implies that $D^k$ is an ideal candidate for the stopping rule: the algorithm can be terminated when $D^k/N$ is below a certain threshold.

{\bf Convergence of primal residuals.}
Figure~\ref{fig:primal} and \ref{fig:primal-1e4} show the trajectory of the primal residual, which is defined as $\sum_i^N \| x_i-z_i \|_2^2$ here.
It reflects how well the constraints $\{ x_i = z_i \}$ are satisfied, and is sometimes called the primal feasibility gap.
For example, if the primal residual is $10^4$ for $N = 10^2$
(or, $10^6$ for $N=10^4$), then
on average each $\| x_i - z_i \|$ is around $10$, which is already
small enough since $x_i$ is in the order of $10^4$. Hence, we conclude that
the constraints are well satisfied after $50$ iterations in both cases.

\begin{figure*}[htbp]
	\begin{minipage}[ht]{0.33\linewidth}
	\centering
	\includegraphics[width=1\linewidth]{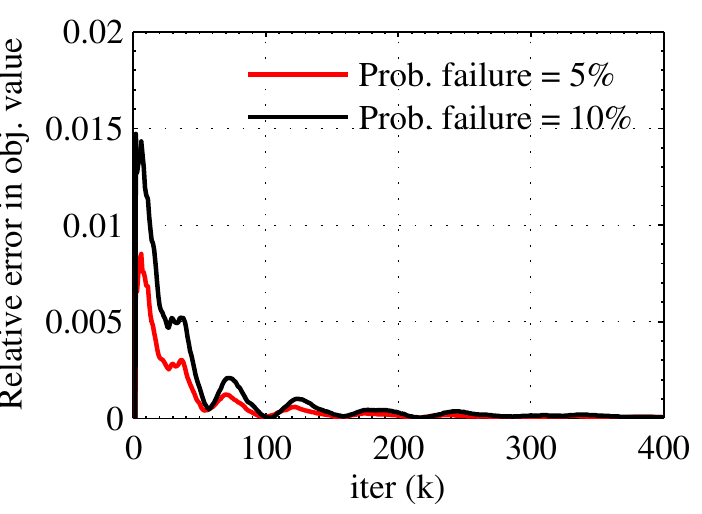}
	\vspace{-5mm}
	\caption{Relative errors in objective value. $N=10^2$.}
	\label{fig:obj-relative}
	\end{minipage}
	\begin{minipage}[ht]{0.33\linewidth}
	\centering
	\includegraphics[width=1\linewidth]{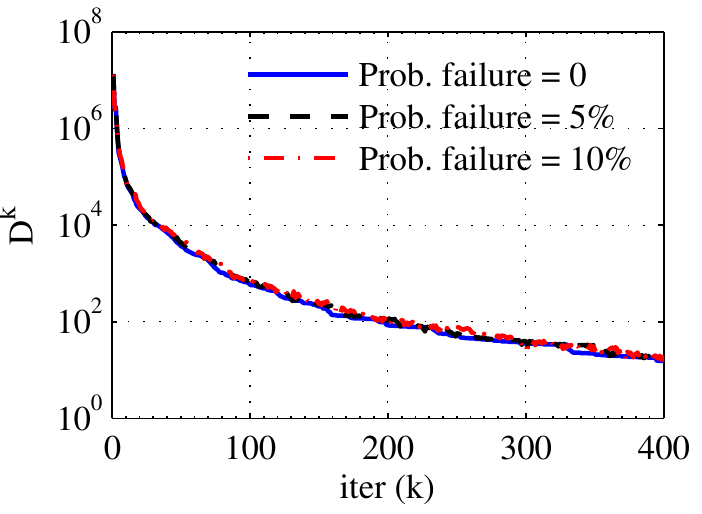}
	\vspace{-5mm}
	\caption{$D^k$. $N=10^2$.}
	\label{fig:Dk-pf}
	\end{minipage}
	\begin{minipage}[ht]{0.33\linewidth}
	\centering
	\includegraphics[width=1\linewidth]{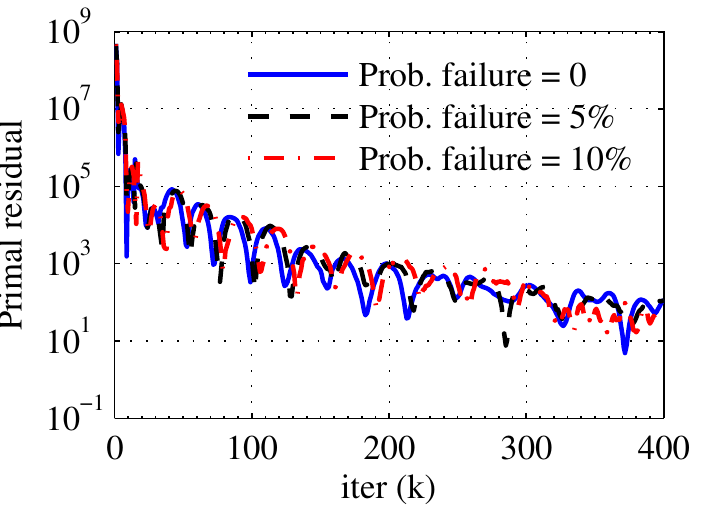}
	\vspace{-5mm}
	\caption{Primal residual. $N=10^2$.}
	\label{fig:primal-pf}
	\end{minipage}
\end{figure*}
\begin{figure*}[htbp]
	\begin{minipage}[ht]{0.33\linewidth}
	\centering
	\includegraphics[width=1\linewidth]{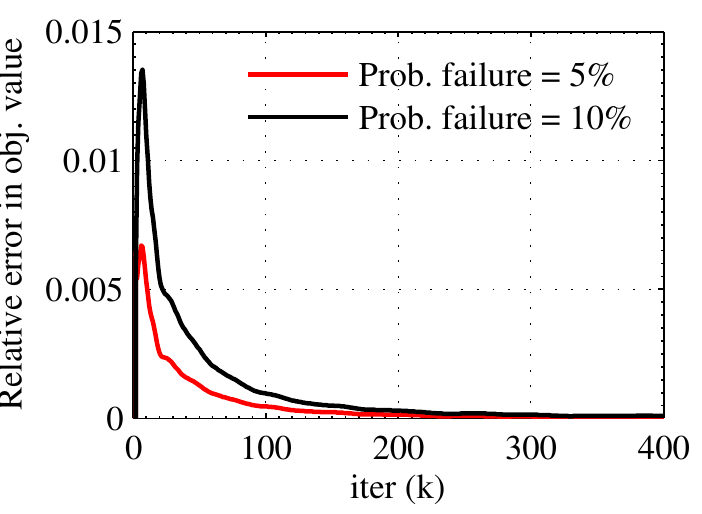}
	\vspace{-5mm}
	\caption{Relative errors in objective value. $N=10^4$.}
	\label{fig:obj-relative-1e4}
	\end{minipage}
	\begin{minipage}[ht]{0.33\linewidth}
	\centering
	\includegraphics[width=1\linewidth]{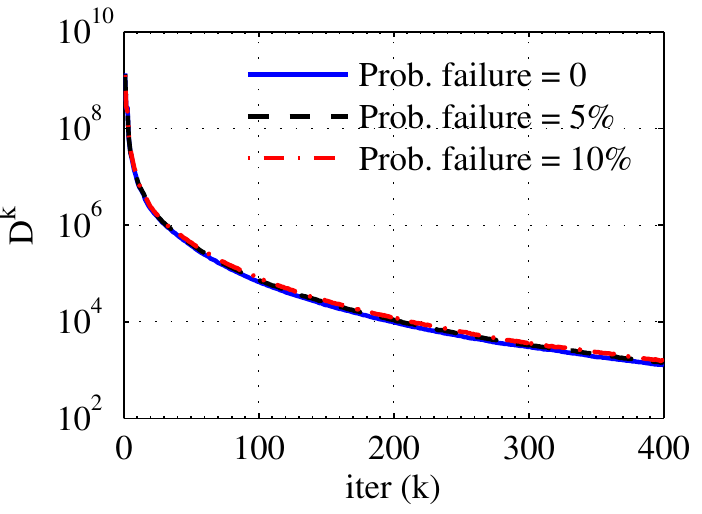}
	\vspace{-5mm}
	\caption{$D^k$. $N=10^4$.}
	\label{fig:Dk-pf-1e4}
	\end{minipage}
	\begin{minipage}[ht]{0.33\linewidth}
	\centering
	\includegraphics[width=1\linewidth]{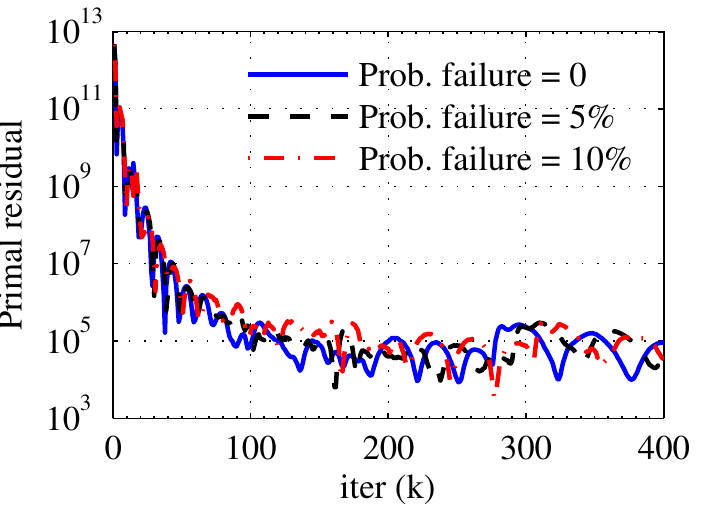}
	\vspace{-5mm}
	\caption{Primal residual. $N=10^4$.}
	\label{fig:primal-pf-1e4}
	\end{minipage}
\end{figure*}

\subsection{Fault-tolerance}

We have observed that our algorithms converge fast to the optimal solution for large-scale problems. Yet, because failures are the norm rather than the exception, fault-tolerance is arguably the most important design objective
for parallel computing frameworks that involve a large number of servers currently \cite{DG04}. A parallel algorithm that is inherently robust against failures in the intermediate steps is highly desirable for practical deployment. To investigate the fault-tolerance of our algorithm, we carry out a new set of simulations where each user fails to update $x_i^k$ with a probability $p$ at each iteration (independent of each other). Whenever a failure happens, user $i$ simply reuses its previous solution by setting $x^{k+1}_i := x^k_i$. 

Figure~\ref{fig:obj-relative}--\ref{fig:primal-pf} plot the convergence with different failure probabilities for $N=10^2$, and Figure~\ref{fig:obj-relative-1e4}--\ref{fig:primal-pf-1e4} for $N=10^4$. Specifically, Figure~\ref{fig:obj-relative} and \ref{fig:obj-relative-1e4} plot the relative error in objective value with failures (i.e. $\textsc{obj\_fail}/\textsc{obj}-1$, where $\textsc{obj\_fail}$ is the objective value with failures, and $\textsc{obj}$ is the objective value when every step is solved correctly). We observe that increasing the failure probability from 5\% to 10\% increases the relative error, causing the solution quality to degrade at the early stage. Yet surprisingly, the impact is very insignificant: The relative error is at most 1.5\%, and ceases to 0 after 100 iterations. In fact, after 50 iterations the relative error is only around 0.2\% for both problem sizes.

Moreover, failures do not affect the convergence of the algorithm at all. This is indicated by the relative error plots, and further illustrated by the overlapping curves in Figure~\ref{fig:Dk-pf}, \ref{fig:primal-pf}, \ref{fig:Dk-pf-1e4}, and \ref{fig:primal-pf-1e4} for $D^k$ and primal residual. 

Thus, we find that our distributed ADMM algorithms are inherently fault-tolerant, with less than 1\% optimality loss and essentially the same convergence speed for up to 10\% failure rate. They are robust enough to handle temporary failures that commonly occur in production systems.

\subsection{Comparison with Dual Decomposition}
\label{sec:sim_dual}
We also simulate the conventional dual decomposition approach with subgradient methods as explained in Sec.~\ref{sec:compare} to solve problem \eqref{opt:basic}. The step size $\rho^k$ is chosen following the commonly accepted diminishing step size rule \cite{BM}, with $\rho^k = 10^{-5}/\sqrt{k}$.

\begin{figure}[htbp]
	\begin{minipage}[ht]{0.49\linewidth}
	\centering
	\includegraphics[width=1\linewidth]{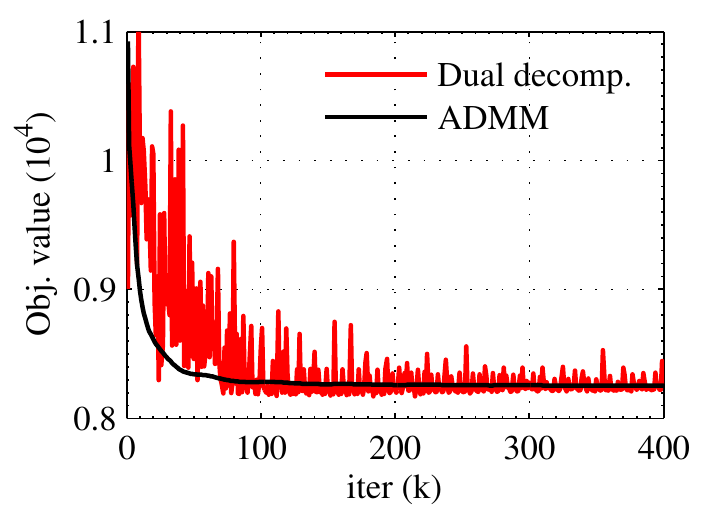}
	\caption{Objective value. $N=10^2$.}
	\label{fig:obj-sm}
	\end{minipage}
	\begin{minipage}[ht]{0.49\linewidth}
	\centering
	\includegraphics[width=1\linewidth]{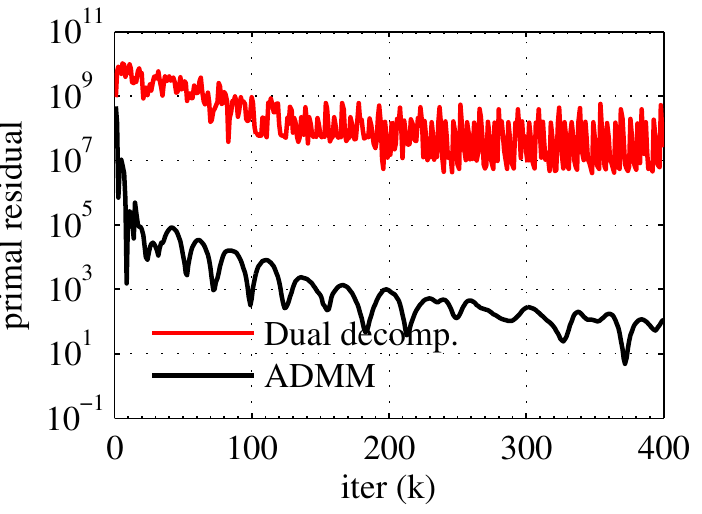}
	\caption{Primal residual. $N=10^2$.}
	\label{fig:primal-sm}
	\end{minipage}
\end{figure}

We plot the trajectory of objective values in Figure~\ref{fig:obj-sm}, and that of primal residuals in Figure~\ref{fig:primal-sm}. Compare to Algorithm~1, dual decomposition yields wildly fluctuating results. Though the objective value decreases to the same level as Algorithm~1 after about 200 iterations, the 
more meaningful primal variables $\{x_i\}$ never converge even after 400 iterations. One can see from Figure~\ref{fig:primal-sm} that the primal residual does not decrease below $10^7$. This implies that the equality constraints $\{x_i=z_i\}$ are not
well-satisfied during the entire course, and the primal variables $\{x_i\}$
still violate the capacity constraints after 400 iterations.

This phenomenon is due to the \emph{oscillation problem} \cite{LS06}
when dual decomposition method is applied to non-strictly convex
objective functions. To mitigate this problem, one can make the objective
function strictly convex by adding a small penalty term, e.g., $\rho_1 \| x \|_2^2 + \rho_2 \| z \|_2^2$). Nevertheless, we found that the primal variables $\{x_i\}$ still converge very slowly after an extensive trial of different
$(\rho_1, \rho_2)$.

To summarize, our simulation results confirm our theoretical analysis, 
demonstrate fast convergence of our algorithms in various settings,
and highlight several additional advantages, especially
the scalability and fault-tolerance.

%% file: related.tex
\section{Related Work}\label{sec:related}

\subsection{Network Utility Maximization}

Network utility maximization (NUM) 
\cite{Bertsekas98,Srikant04}
is closely related to our multi-facility 
resource allocation problem.
A standard technique for solving NUM problems is dual
decomposition. Dual decomposition was first applied to the 
NUM problem in \cite{KMT98}, and has lead to a rich literature
on distributed algorithms for network rate control
\cite{LL99,PC06,CLCD07}
 and new understandings of existing
network protocols \cite{L03}. Despite its popularity, dual decomposition requires a delicate adjustment of the step-size
parameters, which are often difficult to tune.
In addition, dual decomposition requires the utility
functions to be strictly concave and the cost functions
to be strictly convex. Our ADMM-type algorithms overcome these
difficulties, achieving faster convergence under weaker assumptions as discussed
in Sec.~\ref{sec:compare} in detail. 


\subsection{ADMM and Its Variations}

Originally proposed in the 1970s, ADMM has recently received much research attention
and found practical use in many areas, due to its superior empirical performance in solving large-scale
convex optimization problems \cite{BPCP10}. While the convergence of ADMM is well known in the literature
(see, e.g., \cite{BT97,BPCP10}), its rate of convergence has only been established very recently.
\cite{HeY12,HeYuan12} prove rate-$\calO(1/k)$ of convergence for the general case. \cite{DY12} proves rate-$\calO(1/a^k)$ of convergence under
the additional assumptions that the objective function is strongly convex and its
gradient is Lipschitz continuous in at least one block of variables.
These results provide theoretical foundation for our algorithm design and analysis.
ADMM has two important variations: linearized ADMM \cite{HeY12} and multi-block
ADMM \cite{HY12, HL12}. However, they are not particularly suitable for problem
\eqref{opt:form1}, as discussed thoroughly in Section~\ref{sec:compare}. 
In contrast, our ADMM-type algorithms exploit the special structure of problem \eqref{opt:form1},
thereby enjoying a number of unique advantages.


\subsection{Cloud Traffic Management}

Cloud service providers operate two distinct types of WANs: user-facing
WANs and backbone WANs \cite{JKMO13}.
The user-facing WAN connects cloud users and data centers by
peering and exchanging traffic with ISPs.
Through optimized load balancing, this type of networks
can achieve a desired trade-off between performance and cost
\cite{LCBW12,icac13,Henry-sig,QWBG09,LLWL11,XL13,GCWK12}.
The backbone WAN provides
connectivity among data centers for data replication and synchronization.
Rate control and multi-path routing \cite{JKMO13,GHCR13,HKMZ13} 
can significantly increase link utilization and reduce operational costs of the network.
Previous work developed different optimization methods for each application scenario separately, whereas our work
provides a unified framework well suited to a wide range of network
scenarios.

%% file: conclusion.tex
\section{Conclusion}\label{sec:conclusion}

In this work, we have introduced a general framework for studying
various cloud traffic management problems. We have abstracted these
problems as a multi-facility resource allocation problem and
presented two distributed algorithms based on ADMM
that are amenable to
parallel implementation. We have provided the convergence rates of our
algorithms under various scenarios. When the utility functions
are non-strictly concave and the cost functions are non-strictly
convex, our algorithms achieve $\calO(1/k)$ rate of convergence.
When the utility functions are strictly concave \emph{or} the 
cost functions are strictly convex, our algorithms achieve 
$\calO(1/a^k)$ rate of convergence.  

We have shown that, compared to dual decomposition and
other ADMM-type distributed solutions, our algorithms have a number of
unique advantages, such as achieving faster convergence under weaker assumptions,
and enjoying lower computational complexity
and lower message-passing overhead. These advantages are further confirmed by our extensive empirical studies. Moreover, our simulation results demonstrate some additional advantages of our algorithms, including the scalability and
fault-tolerance, which we believe are highly desirable for large-scale cloud systems.

%% file: appendix.tex
\appendix

In this Appendix, we will show that Algorithms~\ref{alg:admm1} and \ref{alg:admm2} are variants of the standard ADMM algorithm.
Let 
$x = (x_1^T, \ldots, x_N^T)^T$, $f(x) = -\sum_{i=1}^N f_i(x_i)$,
$y = (y_1, \ldots, y_n)^T$, and $g(y) = \sum_{j = 1}^n g_j(y_j)$. Then problem \eqref{opt:form1} can be rewritten as:
\begin{align}\label{opt:form2}
\text{minimize} &\quad f(x) + g(y) \\
\text{subject to} &\quad Ax = y \nonumber \\
&\quad x \in \calX, \ y \in \calY, \nonumber
\end{align}
where the matrix $A = [I, \ldots, I]$ ($I$ is the $n \times n$ identity matrix). Although problem \eqref{opt:form2} is in ADMM form, its penalty term 
$(\rho/2)\| \sum_{i=1}^N x_i - y \|_2^2$
violates the separability of the problem.

To address this difficulty, we introduce a set of auxiliary variables $z_i = x_i$, and reformulate problem \eqref{opt:form1} as:
\begin{align}\label{opt:form3}
\text{maximize} &\quad \sum_{i = 1}^N f_i(x_i) - g(\sum_{i = 1}^N z_i) \\
\text{subject to} &\quad \forall i: x_i = z_i \nonumber\\
&\quad \forall i: x_i \in \calX_i; \ \sum_{i=1}^N z_i \in \calY. \nonumber
\end{align}
Now, the new penalty term is $(\rho/2) \sum_{i=1}^N 
\|x_i - z_i\|_2^2$, which preserves separability.

Applying the scaled form of ADMM to problem \eqref{opt:form3},
we obtain the following iterations:
 \begin{align*}
  	x_i^{k+1} &:= \argmin_{x_i \in \calX_i} \left( -f_i(x_i)
  	+ (\rho/2) \|x_i - z_i^{k} + v_i^k \|_2^2 \right)\\
  	z^{k+1} &:= \argmin_{(\sum_{i} z_i) \in \calY} \left( g(\sum_{i = 1}^N z_i) + (\rho/2)
  	\sum_{i = 1}^N \| z_i - x_i^{k+1} - v_i^k \|_2^2 \right) \\
  	v_i^{k+1} &:= v_i^k + x_i^{k+1} - z_i^{k+1}.
 \end{align*}
We will show that the above iterations are equivalent
to Algorithm~\ref{alg:admm1}. The key observation is that
the dual variables $v_i^k$ are equal for all the users,
i.e., $\forall i: v_i^k = v^k$, as shown in \cite[Chapter~7]{BPCP10}.

Let $u^k \triangleq \sum_{i=1}^N v_i^k = N v^k$ and
$y^k \triangleq \sum_{i=1}^N z_i^k$. Then, the dual update
can be rewritten as 
\[
u^{k+1} := u^k + \sum_i x_i^{k+1} - y^{k+1},
\]
which is exactly the dual update in Algorithm~\ref{alg:admm1}.

Substituting $v^k = v^{k-1} + x_i^k - z_i^k$ 
and 
\[
v^k = v^{k-1} + (1/N) \left( \sum_i x_i^k - y^k \right)
\]
in the $x$-update gives
\[
x_i^{k+1} := \argmin_{x_i \in \calX_i} \left( -f_i(x_i)
  	+ (\rho/2) \|x_i - x_i^k + d^k \|_2^2 \right),
\]
which is exactly the $x$-update in Algorithm~\ref{alg:admm1}. 

Finally, substituting 
\[
z_i^{k+1} - x_i^{k+1} - v_i^k = -v^{k+1} = (1/N) \left( y^{k+1} 
- \sum_i x_i^{k+1} - u^k \right)
\]
in the $z$-update gives
\[
y^{k+1} := \argmin_{y \in \calY} g(y) + (\rho/2N)
\| y - \sum_i x_i^{k+1} - u^k \|_2^2,
\]
which is precisely the $y$-update in Algorithm~\ref{alg:admm1}.
Hence, Algorithm~\ref{alg:admm1} is indeed a variant of the
standard ADMM algorithm. 

Similarly, we can show that 
Algorithm~\ref{alg:admm2} is 
equivalent to the following iterations:
 \begin{align*}
  	z^{k+1} &:= \argmin_{(\sum_{i} z_i) \in \calY} \left( g(\sum_{i = 1}^N z_i) + (\rho/2)
  	\sum_{i = 1}^N \| z_i - x_i^{k} - v_i^k \|_2^2 \right)\\
 	x_i^{k+1} &:= \argmin_{x_i \in \calX_i} \left( -f_i(x_i)
 	+ (\rho/2) \|x_i - z_i^{k+1} + v_i^k \|_2^2 \right)\\
  	v_i^{k+1} &:= v_i^k + x_i^{k+1} - z_i^{k+1}
 \end{align*}
which can be viewed as the scaled form of ADMM 
with the order of $x$-update and $z$-update switched.